# Good Practice Guide on the electrical characterization of graphene using non-contact and high-throughput methods

**Edited by:**
Alexandra Fabricius (VDE)
Alessandro Cultrera (INRIM)
Alessandro Catanzaro (NPL)

**Contributions from:**
Alessandro Catanzaro (NPL)
Nathaniel J. Huang (NPL)
Christos Melios (NPL)
Ling Hao (NPL)
John Gallop (NPL)
Israel Arnedo (das-Nano)
David Etayo (das-Nano)
Elena Taboada (das-Nano)
Alessandro Cultrera (INRIM)
Olga Kazakova (NPL)

**Affiliations:**
VDE, Association for Electrical, Electronic & Information Technologies, Frankfurt am Main, Germany.
INRIM, Istituto Nazionale di Ricerca Metrologica, Turin, Italy.
NPL, National Physical Laboratory, Teddington, United Kingdom.
das-Nano, Tajonar, Navarra, Spain.

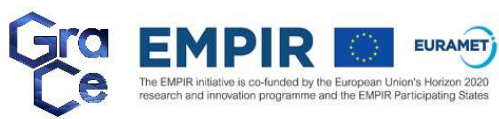

# Index

# Guide information

**What is it about?**

This guide provides protocols for determining the electrical properties of graphene sheets, on insulating substrates, using novel methodologies for non-contact and high-throughput methods: the presented methods do not imply any physical contact between the sample surface and the probing medium.

Depending on the methodology the properties that can be measured are the electrical sheet resistance, the concentration and mobility of electrical charge carriers. Each protocol gives advice to understand the measurement principle, how to implement it to perform reliable measurements traceable to the International System of units, and hints to express the corresponding measurement uncertainty.

**Who is it for?**

The guide is for producers and users of graphene who need to understand how to measure the electrical properties of graphene. Such information is essential in a host of technology application areas where graphene may be used.

**What is its purpose?**

This guide provides a detailed description of how to determine the key electrical properties of graphene, so that the graphene community can adopt a common, metrological approach that allows the comparison of commercially available graphene materials. The guide is intended to form a bedrock for future interlaboratory comparisons and support the direct collaboration with international standardisation organisations.

**What is the pre-required knowledge?**

The guide is for users in research and industry who have experience with and access to the advanced techniques described herein. It is targeted at analytical scientists and professionals who have a Bachelor's degree in science.

# Overview

## Introduction

This guide is a deliverable of the Joint Research Project 16NRM01 GRACE, *Developing electrical characterisation methods for future graphene electronics*. The project belongs to the European Metrology Programme for Innovation and Research (EMPIR). GRACE is framed within the Normative targeted program, and its overall goal is (1) the development of validated protocols for the measurement of the electrical properties of graphene, and their implementation in order to achieve accurate and fast-throughput measurement of graphene; and (2) the collaboration with international standardisation committees in order to initiate and develop dedicated documentary standards for the electrical characterisation of graphene. The adoption of this GPGs and, when published, the corresponding standards, will allow industry to perform accurate measurements of the electrical properties of graphene and thereby provide customers with reliable and comparable specifications of graphene as an industrial product.

## Scope

The electrical characterisation of graphene, either in plane sheets or in properly geometrised form can be approached using non-contact methods already employed for thin film materials. The extraordinary thinness (and, correspondingly, the volume) of graphene, however, makes the proper application of these methods difficult. The electrical properties of interest (sheet electrical resistivity/conductivity, concentration and mobility of charge carriers) must be indirectly derived from the measurement outcome by geometrical and electrical modelling; the assumptions behind such models (e.g., uniformity and isotropy, effective value of the applied fields, etc.) require careful consideration. The traceability of the measurement to the International System of units and a proper expression of measurement uncertainty is an issue.

This guide focuses on non-contact and high-throughput methods, that are methods where the graphene sample surface is not physically contacted with any metallic electrodes at any stage. A companion guide about contact methods is also available.

The methods discussed are:

- Measurement of surface potential and work function using Scanning Kelvin Probe Microscopy (SKPM);
- Measurement of sheet resistance by Microwave Resonant Cavity;
- Measurement of sheet resistance by Terahertz time-domain spectroscopy (THz-TDS);

For each method, a corresponding measurement protocol is discussed, which describes:

- The measurement principle;
- Sample requirements and preparation;
- A description of the measurement equipment / apparatus;
- Calibration standards and ways to achieve a traceable measurement;

- Environmental conditions to be considered;
- A detailed measurement procedure, with specific hints to achieve a reliable measurement;
- modeling and data analysis to determine the electrical property of interest;
- considerations about the expression of measurement uncertainty.

# Terms and Definitions

The International Standardisation Organisations ISO (International Organization for Standardisation) and IEC (International Electrotechnical Commission) have published many standards in the area of measurement and characterisation of nanomaterials, particularly referring to some of the techniques detailed here. For further information the IEC/TC 113 'Nanotechnology in electrotechnical products and systems', ISO/TC 229 'Nanotechnologies', ISO/TC24/SC4 'Particle characterization' and ISO/TC 201 'Surface Chemical Analysis' websites should be examined, which list both the published and under-development standards in this area.

Furthermore, ISO and IEC maintain terminological databases for use in standardization:

- IEC Electropedia: available at http://www.electropedia.org/
- ISO Online browsing platform: available at http://www.iso.org/obp

In particular, the terms and definitions from "ISO TS 80004-13: Nanotechnologies -- Vocabulary -- Part 13: Graphene and related two-dimensional (2D) materials" apply here and should be referred to.

For the purposes of this document, the following terms and definitions apply:

**Graphene; graphene layer; single layer graphene; monolayer graphene**
single layer of carbon atoms with each atom bound to three neighbours in a honeycomb structure

*Note 1 to entry: It is an important building block of many carbon nano-objects.*

*Note 2 to entry: As graphene is a single layer, it is also sometimes called monolayer graphene or single layer graphene and abbreviated as 1LG to distinguish it from bilayer graphene (2LG) and few-layered graphene (FLG).*

*Note 3 to entry: Graphene has edges and can have defects and grain boundaries where the bonding is disrupted.*

[Source: ISO/TS 80004-13]

**Bilayer graphene (2LG)**
two-dimensional material consisting of two well-defined stacked graphene layers

*Note 1 to entry: If the stacking registry is known, it can be specified separately, for example, as "Bernal stacked bilayer graphene".*

[Source: ISO/TS 80004-13]

**Few-layer graphene (FLG)**
two-dimensional material consisting of three to ten well-defined stacked graphene layers

[Source: ISO/TS 80004-13]

**Electrical conductivity, *σ***
the SI unit of measure of σ in graphene and 2D materials is the siemens (S)

**Sheet resistance, $R_S$**
the electrical resistance of a conductor with a square shape (width equal to length) and thickness significantly smaller than the lateral dimensions (thickness << width, length)

*Note 1 to entry: The SI unit of sheet resistance is ohms (Ω). However, for the purpose of this procedure, expressed as ohm per square (Ω/sq).*

*Note 2 to entry: We define here a "sheet resistivity" as the intrinsic, local property of two-dimensional conductor, a two-dimensional equivalent to three-dimensional resistivity. For an electrically uniform thin film material, the sheet resistance and sheet resistivity are identical.*

[Source: adapted from IEC/TS 61836:2007 Ed. 2.0]

**Drift mobility of a charge carrier, *μ***
the quotient of the modulus of the mean velocity of the charge carriers in the direction of an electric field by the modulus of the field strength. The drift mobility is expresses in the International system as $cm^2*s/V$

*Note 1 to entry: The (drift) mobility is here considered to be the fundamental, intrinsic (local) property. The Hall and field effect mobility are then the extrinsic (sample) electrical measurements, carried out to determine the intrinsic mobility.*

*Note 2 to entry: The (drift) mobility for electrons and holes can be very different, depending on the residual doping and scattering mechanisms for the given sample.*

[Source: adapted from IEV 521-02-58]

**Surface potential, $U_{CPD}$**
is the electrostatic potential near the material surface

**Work function, *Φ***
is the energy required to remove an electron from the surface of the target material to a point in vacuum outside the solid surface, where the potential of the material can be considered negligible

*Note 1 to entry. The work function is a property of the surface of the material and not related to the bulk phase.*

**Hall mobility, $\mu_H$**
the product of the Hall coefficient and the sheet conductance, as obtained from a Hall measurement

*Note 1 to entry. Similar considerations apply to Hall mobility measurements as to Field effect measurements, with respect to contact resistance.*

*Note 2 to entry. Even for a uniform sample, the Hall mobility differs with the drift velocity by a scattering factor: r = µH / µ. The Hall scattering factor is not generally known in graphene, but is calculated for epitaxial graphene [4] to be of order 0.8-0.95 depending on temperature.*

[Source: adapted from IEV 521-09-04]

**Mean free path**
the product of the momentum relaxation time and the Fermi velocity

*Note 1 to entry: The definitions in this document apply only to conductors which are diffusive, i.e. have lateral (parallel to the plane) dimensions larger than relevant transport lengths, the mean free path (momentum relaxation length) as well as the de Broglie (Fermi) wavelength (LF). Sheet resistivity and mobility is not defined for conductors smaller than the mean free path.*

*Note 2 to entry: CVD graphene on $SiO_2$ have mean free paths in the range 1-200 nm at room temperature, while CVD graphene encapsulated in hexagonal boron nitride, can have mean free paths up to 2 µm at room temperature. [SOURCE: adapted from IEV 521-09-04]*

*Note 3 to entry: The mean free path discussed in diffusive transport should be defined, as here, in terms of the relaxation of carriers rather than the actual time between discrete scattering events. Only in the special case of elastic, isotropic (momentum-randomising) scattering events, the distance between scattering events and the momentum relaxation length (=mean free path) are identical.*

**Diffusive conductor**
a conductor where the dimensions in which electron transport takes place are significantly larger than the mean free path

**Two-dimensional conductor**
a conducting object or material where the motion of the carriers is fully confined in a direction while unrestricted on the two other directions perpendicular to the first one (i.e. no quantum degrees of freedom in z-direction)

**Sheet conductance, $G_s$**
the inverse of sheet resistance, $G_s = 1/R_s$

**Non-contact microwave method, Microwave cavity method, MW Cavity method**
the method to measure average surface conductivity or equivalently sheet resistivity by resonant cavity involves monitoring the resonant frequency shift and change in the quality factor before and after insertion of the specimen into the cavity in a quantitative correlation with the specimen surface area

*Note 1 to entry: The method is fast and non-contacting.*

[Source: adapted from ISO 80004-13, 5.3.9]

**Terahertz time domain spectroscopy**
the method to measure the complex-valued dielectric function or conductivity of a material in the THz frequency range (typically 0.1 - 5 THz) by measurement of the temporal shape of a picosecond-duration electromagnetic pulse either reflected from or transmitted through the sample

*Note 1 to entry: The amplitude and phase of the frequency components of the signal are compared to those of a reference signal, and can be related to the complex refractive index, permittivity, or conductivity of the sample.*

**Time-domain Signal-to-Noise Ratio, TD-SNR**
the square of the ratio between the measured peak-to-peak electric field signal and the RMS value of the measured noise over the same time window length with the THz beam blocked. This SNR value is expressed in electric field

**Frequency-domain Signal-to-noise Ratio, FD-SNR**
the frequency-dependent quantity defined as the square of the ratio between the maximum of the measured amplitude spectrum and the RMS value of the corresponding amplitude noise spectrum measured with the terahertz beam blocked

**Frequency-domain Dynamic Range, FD-DR**
the frequency-dependent quantity defined as the ratio between spectral amplitude of the THz electric field signal and the spectral amplitude of the noise, measured with the THz beam blocked

**Spot-size** (for THz-TDS section)
the size of the terahertz spot on the sample

*Note 1 to entry: The THz beam is typically of very broadband nature. Therefore, the spot-size of a THz beam can either be measured at a specific frequency within its bandwidth, or as an average value by a superposition of spot sizes at all frequencies, weighted by their spectral amplitude. The spot size is typically given as FWHM (full width at half maximum) of the spatial field distribution.*

# 1 Measurement of surface potential and work function using Scanning Kelvin Probe Microscopy

This part of the protocol establishes a standardized method to determine the surface potential and work function by the Scanning Kelvin Probe Microscopy method.

Scanning Kelvin Probe Microscopy (SKPM, often called also as Kelvin probe force microscopy, KPFM) is a contactless atomic force microscopy technique capable of measuring the surface potential and therefore the local work function of graphene samples. Frequency modulated-SKPM (FM-SKPM) detects the force gradient using the frequency shift to calculate the contact potential difference (often called surface potential) between the tip and the sample. By calibrating the metal-coated AFM probe, the absolute work function of the sample can be measured.

## 1.1 General

### 1.1.1 Measurement principle

Atomic force microscopy (AFM) is a scanning probe microscopy (SPM) technique that provides high-resolution topographic information through the van der Waals interactions between the tip and the sample (attractive-repulsive forces). Usually, a pyramidal tip is attached at the end of a mechanically oscillating ($f_0$ ~ 50-350 kHz) cantilever, which is typically made of silicon or silicon nitride. The mechanical oscillations are induced using a piezoelectric material, and the whole tip-piezoelectric system is placed in a piezoelectric tube dedicated to executing x, y (~100 μm) and z (~10 μm) movements controlled by a feedback loop. The feedback loop uses a laser that reflects from the top of the cantilever and onto a four-quadrant photodetector to maintain the deflection set-point. By recording the feedback height (z) and the x, y movements during the scan, a 3D map is generated.

There are two main modes for operating an AFM. The simplest one is the contact mode. In this mode, the tip is coming in contact with the surface, and by scanning, the feedback loop maintains a small set-point which is recorded to build the 3D map. The second mode is the tapping mode. Here, the cantilever is in the lifted position and is oscillating at the resonant mechanical oscillation frequency. The forces between sample and tip will then shift the resonance frequency, the oscillation amplitude ($A_{osc}$) and the oscillation phase ($\varphi$) of the cantilever. By adjusting these parameters through the feedback loop and recording them, a topography map is constructed.

SKPM is a tapping mode technique, which can either be used in a single pass, where at the same time topography and surface potential ($U_{CPD}$) are being recorded, or in double pass, where the first pass records topography and the second, in lift mode, records the surface potential by tracing the first pass. SKPM is measuring the contact potential difference (CPD) when a conductive tip is in proximity with the sample. In this case:

$$U_{\text{CPD}} = \frac{\Phi_{\text{S}} - \Phi_{\text{t}}}{-e} \tag{1.1}$$

where $\Phi_S$ and $\Phi_t$ are the work functions of sample and tip, respectively. This is because of the difference in Fermi levels of the two materials. When the two Fermi levels are aligned, the system is in equilibrium. When an external bias voltage is applied ($U_{\text{probe}}$), the $U_{\text{CPD}}$ will be nullified.

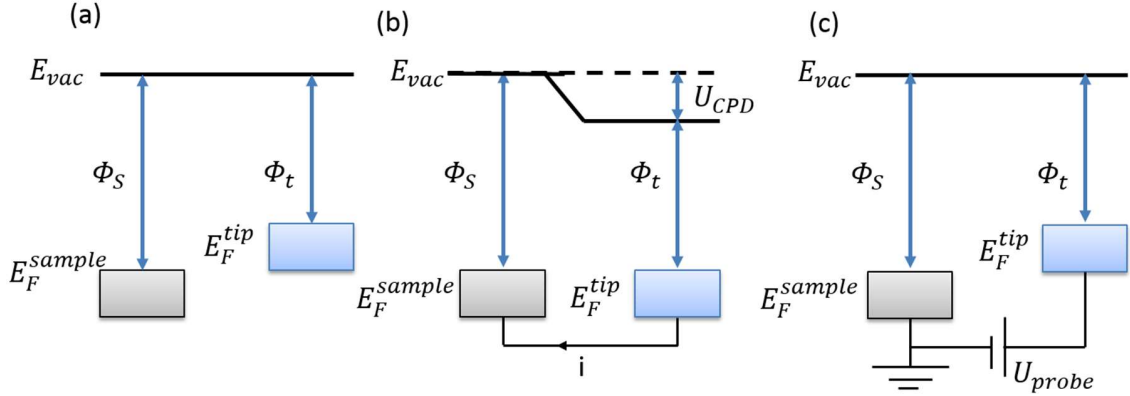

**Figure 1.1: (a) Sample and tip are at some distance, (b) sample and tip are electrically connected and $U_{CPD}$ is generated and (c) $U_{probe}$ is applied to nullify the $U_{CPD}$.**

In the case of the double pass, the mechanical excitation of the cantilever is deactivated and an AC voltage (known as $V_{\text{mod}}$) is applied. An oscillatory force is generated by the AC voltage when $U_{\text{probe}} \neq U_{\text{CPD}} + U_{\text{induced}}$. The oscillation at the first ($F_\omega$) and second ($F_{2\omega}$) harmonic is described by:

$$F_\omega = -\frac{1}{2}\frac{dC}{dz}\left[\left(U_{\text{probe}} - (U_{\text{CPD}} + U_{\text{induced}})\right) + V_{\text{mod}}\sin(\omega t)\right]^2 \quad (2.1)$$

and:

$$F_{2\omega} = -\frac{1}{4}\frac{dC}{dz}V_{\text{mod}}^2 \cos(2\omega t) \quad (2.2)$$

where $\omega = 2\pi f_0$. Frequency-modulated SKPM (FM-SKPM) uses the force gradient $dF_\omega/dz$ to calculate $U_{\text{CPD}}$. This is done by measuring a mechanical resonant frequency shift of the cantilever either in single or double pass. In single-pass FM-SKPM, the cantilever is oscillating at its mechanical resonant frequency $f_0 \approx 300$ kHz, with a much lower AC frequency voltage $f_{\text{mod}} \approx 3$ kHz also applied, to induce a frequency shift of $f_0 \pm f_{\text{mod}}$. The side lobes (monitored by a PID feedback loop) generated by this shift are minimized by applying a DC compensation voltage. By measuring this DC voltage at each pixel, a surface potential map [i.e. contact potential difference ($U_{\text{CPD}}$)] is constructed. Because in FM-SKPM the force gradient is being detected, a spatial resolution of <20 nm can be achieved, which is limited only by the tip apex diameter [1]. In FM-KPFM, the oscillation amplitude ($A_{\text{osc}}$) is described by:

$$A_{\text{osc}}(f_0 \pm f_{\text{mod}}) \approx f_0\left(1 - \frac{1}{2k}\frac{dF_\omega}{dz}\right) \quad (2.3)$$

To obtain the $U_{\text{CPD}}$ and build the map, the feedback loop minimises the frequency shift by applying $U_{\text{probe}}$. In the case where double pass is used, first the topography is recorded and in the second pass the cantilever traces the measured topography to measure surface potential [1]–[4]. The working principle of FM-KPFM is shown in Figure 1.2.

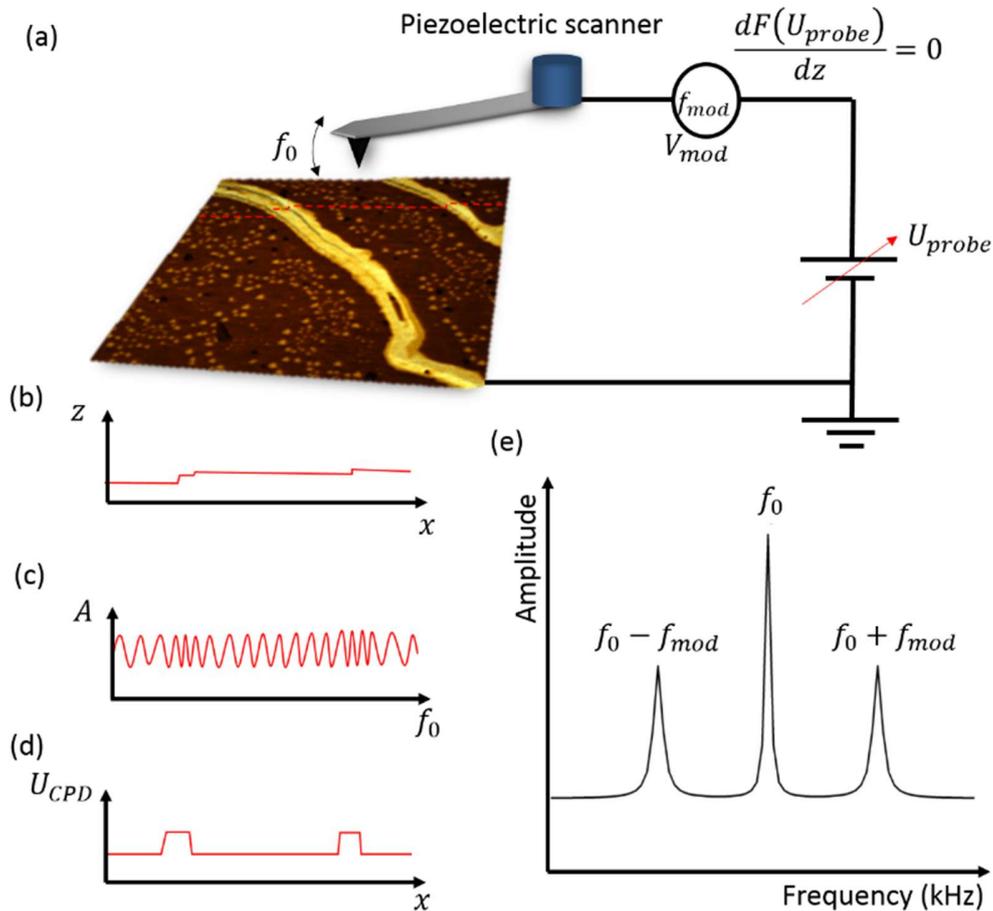

**Figure 1.2: The working principle of FM-FPFM**

Figure 1.2 shows a (a) FM-SKPM operation in single-pass tapping mode to measure the sample topography (b). At the same time an AC voltage with much lower frequency ($f_{\mathrm{mod}}$) is applied to the tip, which induces $f_0 \pm f_{\mathrm{mod}}$ side lobes (d). By minimizing these side lobes using a feedback loop, the surface potential ($U_{\mathrm{CPD}}$) is measured. The side-lobes at the frequency spectrum of the cantilever are shown in (e).

### 1.1.2 Sample preparation method

Most importantly, the samples need to be properly grounded prior any SKPM measurements, to avoid charging and therefore incorrect surface potential measurements. The resistance between the sample and ground point needs to be checked prior any measurements.

Despite SKPM being able to measure any conductive sample, when a single-atomic layer such as graphene is measured, surface contamination may lead to incorrect results. When graphene is transferred on a target substrate, the top surface is covered with polymer residues. In our measurements, the polymer contaminant has been removed prior to the SKPM measurement trough a contact AFM scan, which pushed the contaminants away from the region of interest.

### 1.1.3 Description of measurement equipment / apparatus

Figure 1.3 shows the schematic diagram of SKPM setup. The lower part of the schematic illustrates the components for topography measurements and the upper part the components for obtaining a surface potential map. These include the SKPM controller and lock-in amplifier. An AC voltage ($V_{ac}$) is being applied on the tip using the Lock-in reference signal output (OSC out). In FM configuration, the frequency shift signal ($\Delta f$) is split into two paths; the first one is used by the z regulator for topographic imaging, and the other is fed into the lock-in amplifier, which extracts the signal with the same frequency as $V_{ac}$ and feeds the signal into the SKPM controller. The SKPM controller maintains feedback to nullify the lock-in output signal, by applying $V_{dc}$ to the tip. In AM mode, $V_{ac}$ with the same frequency as the second resonant peak of tip oscillation is applied to the AFM tip to excite the tip with electrical force. The amplitude of tip oscillation has two components; low frequency (the first resonance peak) tuned by mechanical oscillation and high frequency (the second resonance peak) tuned by $V_{ac}$. A band-pass filter filters the low and high frequency signals. The low frequency signal is used for topography regulation. The high frequency signal feeds directly into the lock-in amplifier. The $V_{ac}$ controller measures the surface potential (CPD) using the second resonance frequency component [2].

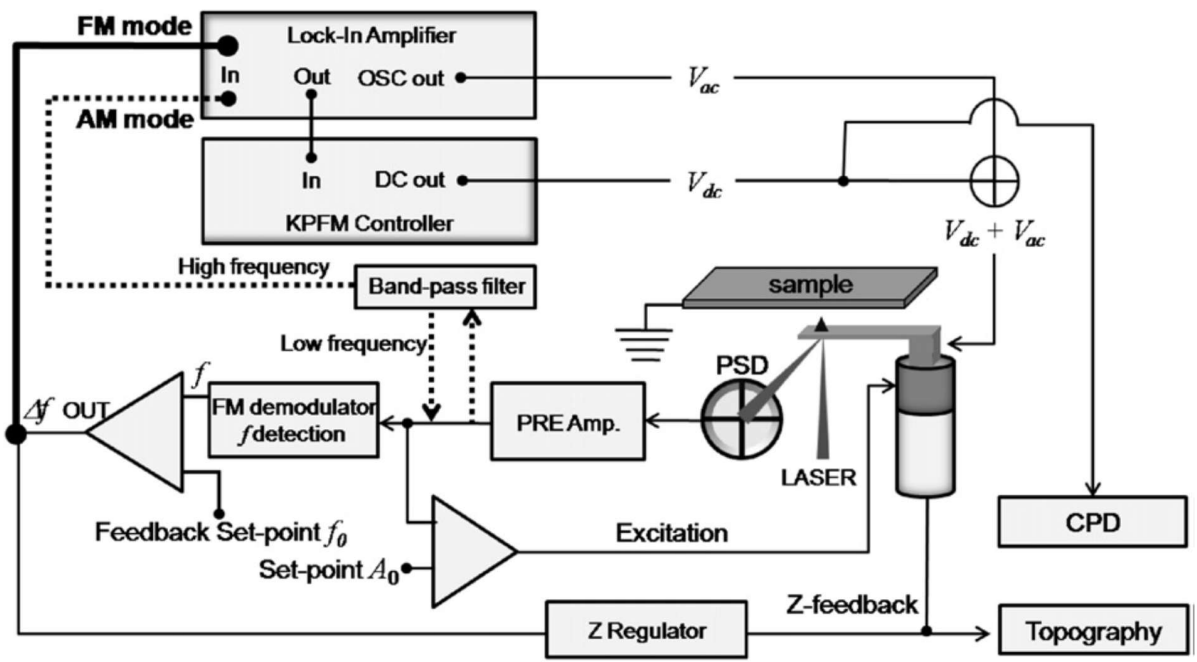

**Figure 1.3: Schematic diagram of SKPM system showing AM (dashed line) and FM (bold line) mode. Lower part of the diagram is an FM mode AFM system for topography imaging and upper part is a SKPM system for surface potential measurement. Reproduced from Ref. [2]**

### 1.1.4 Supporting materials

The preparation of atomically sharp and conducting AFM tips is critical for high-resolution KPFM. An AFM tip can be prepared for high-resolution KPFM using any of several methods. The most common method is to take a commercially available heavily doped Si cantilever and use heat treatment followed by $Ar^+$ bombardment to remove the native oxide layer and other contaminants. These tips routinely produce high- resolution images, but are more likely to pick up surface atoms, altering the work function of the tip [2].

### 1.1.5 Calibration standards

**Scanner X-Y calibration:** Use calibration grids with known period and roughness (<10 nm).

**Scanner Z calibration:** Use calibration grids with known period, height and roughness (<10 nm).

**SKPM tip work function calibration:** Typically, any conductive material with known and stable work function is suitable for calibrating the SKPM tips in order to extract the absolute work function of graphene. However, the most widely used calibration sample is highly oriented pyrolytic graphite (HOPG).

### 1.1.6 Ambient conditions during measurement

Being a carbon monolayer, graphene electrical properties are strongly affected by ambient conditions, in particular, the humidity present in air [2]. The ambient conditions during the whole duration of the measurement must be monitored and recorded. Ambient conditions must be T = (23 ± 1)°C, RH = (50 ± 4) % as specified in table 1 of IEC 60068-1. However, measurements in vacuum offer significant advantage in both spatial and electrical resolution.

## 1.2 Measurement procedure

### 1.2.1 Calibration of measurement equipment

**Scanner X-Y calibration:** It is recommended that calibration takes place at least every 6 months. The calibration procedure will be explained:

1) Take a scan of the test grating. The following conditions should be met:
   - The periodicity direction of the test grating should be oriented along the calibration direction with accuracy not worse than several degrees.
   - For the scanner with sensors, the XY feedback loop should be closed.
   - Scanning goes over the maximum scan area.
2) Use an image processing software to calculate the FFT of the collected image and compare the results to the calibration grid data.
3) If needed, apply a calibration factor in the AFM software.
4) Repeat the scan to ensure the calibration was applied correctly.

**Scanner Z calibration:** It is recommended that calibration takes place at least every 6 months. The calibration procedure will be explained:

1) Take a scan of the test grating. The following conditions should be met:
   - Direction of fast scanning should be close to the periodicity direction of the grating.
   - The scan area should cover at least several periods of the grating. For correct processing of acquired data, numbers of scan points along each direction should be multiple of 2*n*. Recommended size of the scan is at least 512 × 512.
2) Using an image processing software, rotate the image, so the grid lines are perpendicular to the scan direction. Calculate the mean grid height.
3) If needed, apply a calibration factor in the AFM software.
4) Repeat the scan to ensure the calibration was applied correctly.

**SKPM tip work function calibration:** The calibration method is presented in figure 1.4. Firstly, freshly cleaved HOPG is measured using ultra-violet photoelectron spectroscopy. Therefore, the work function of HOPG is calculated using $\Phi_{\text{HOPG}} = h\nu - x = 4.48$ eV, where $h\nu = 21.22$ eV is the photon energy and $x$ is the energy difference between the Fermi edge and cut-off. Then the same HOPG reference sample is measured using FM-SKPM, and the tip is calibrated using $\Phi_{\text{Tip}} = \Phi_{\text{HOPG}} + eU_{\text{CPD}}^{\text{HOPG}}$, where $U_{\text{CPD}}^{\text{HOPG}}$ is the surface potential measured using FM-SKPM. Subsequently to the calibration, the graphene sample is measured, and the work function can be calculated using $\Phi_{\text{Gr}} = \Phi_{\text{Tip}} - eU_{\text{CPD}}^{\text{Gr}}$. A good practice is to check the calibration of the tip after any measurements to ensure no drift or errors.

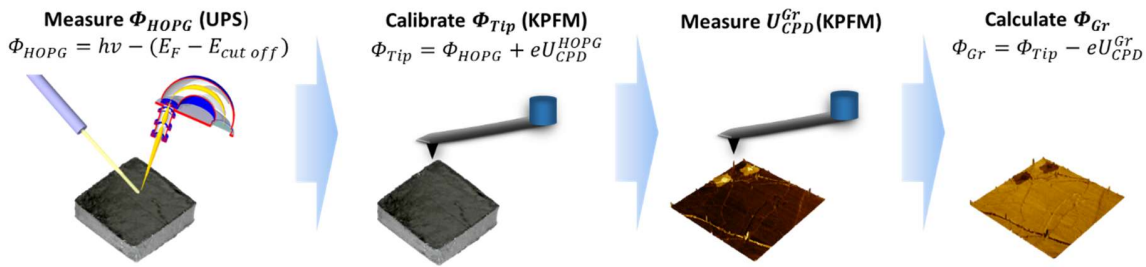

**Figure 1.4: - Calibration and measurement procedure of local work function of graphene samples.**

## 1.2.2 Detailed protocol of the measurement procedure

The following procedure must be followed in order to obtain a surface potential measurement of graphene:

1. Turn on equipment and ensure enough time has passed for the electronics to warm up (~30 minutes).

2. Load AFM tip in the probe holder.

3. Align the AFM laser. Use the diffraction pattern of the AFM cantilever or the optical camera (if available) to align the laser at the end of the AFM cantilever.

4. Zero the PSD sensor (both vertical and horizontal photodiodes).

5. Load the graphene sample on the sample holder and ensure it is grounded. If calibrated SKPM measurements are required, place the calibration standard next to the sample and ground.

6. Collect the amplitude-frequency response (AFR) of cantilever oscillation in the predefined frequency range.

7. Using this AFR, the resonance peak is located and its frequency (*f*) is determined, then the AFR range is corrected. Further, the AFR is taken over the corrected range and refined resonance frequency is found.

8. Adjust the output amplitude and the gain of the cantilever oscillations at its predefined level. The phase of the generator output is tuned so as the phase shift between the detected signal and the reference signal is at its predefined level.

9. Focus the optical camera on the AFM cantilever.

10. Manually bring the sample and cantilever close (take caution to avoid crashing the tip on the sample surface).

11. Define the tapping set-point of the measurement and approach the surface with caution.

12. When the AFM tip is in contact with the sample make the appropriate fine adjustments on the set-point and gain.

13. Select the scan location and adjust the scan area, resolution and rate. When choosing a scanning rate, the user must consider the scan size to avoid excessive scan speeds.

14. Select the single-pass SKPM mode and adjust the AC voltage amplitude, frequency and adjust the phase. Typical values are $V_{ac} = 2 - 8\,V$ at $f_{ac} = 2 - 8\,KHz$.

15. Begin scan and adjust both AFM and SKPM gains so that the trace and re-trace of the scan match each other to the closest extend, while maintaining reasonable signal-to-noise ratio. These adjustments depend highly on the sample and AFM tip used. In some cases, fine adjustments on the phase may needed.

### 1.2.3 Measurement accuracy

In general, the resolution and accuracy of FM-SKPM may depend on several factors, however a mathematical expression of uncertainty is still not available. In brief FM-SKPM demonstrated:

– Excellent agreement with macroscopic UPS measurements.
– The FM method becomes accurate for objects larger than the tip radius. Therefore, the FM method is best used with the sharpest tips and might even achieve atomic contrast with accurate values.
– The FM method does not show any variation of the detected surface potential within a tip-sample separation of 30 nm. Therefore, artefacts due to variations in topography can be excluded in FM-SKPM.
– FM-SKPM works properly for stiff cantilevers, clearly separates topography and Kelvin probe feed- back, and allows the use of small modulation voltages.

Nevertheless, for work function calibrated measurements, most of the uncertainty is introduced due to the calibration of the AFM tip, which included both the uncertainty of the UPS and SKPM measurements on the reference sample. However, this is not the case for qualitative surface potential measurements.

## 1.3 Data analysis / interpretation of results

For data analysis, the user should use any available SPM image processing software available. The user should apply any appropriate flattening and noise filters on the topography map, however the surface potential map should remain unaltered (if any processing takes place, the raw data values will change).

## 1.4 Results to be reported

### 1.4.1 General

The results of the measurement shall be documented in a measurement report, including the date and time of the measurement as well as the name and signature of the person responsible for the accuracy of the report.

### 1.4.2 Sample identification

The report shall contain all information to trace back the history of the sample, such as:

- Supplier;
- Trade name;
- Traceability method: Batch number, serial number, others;
- ID Number;
- Manufacturing date;
- Specification;
- Material type;
- Substrate;
- Sample geometry: Technical drawing or similar.

### 1.4.3 Test conditions

The ambient conditions during the test:

- Atmosphere;
- Temperature range: 24 °C < T < 26 °C;
- Range of relative humidity: 45 % < RH < 55 %.

### 1.4.4 Measurement specific information

Documentation on the applied setup should comprise information on the applied measurement device (type, manufacturer). Especially:

- Calibration status;
- Measurement configuration and scan parameters;
- Sample thickness (including substrate of the layer to be investigated);
- Sample size;
- Sampling plan for single point mode.

### 1.4.5 Test results

The test results shall be documented and comprise the following information:

- Date and time of the measurement;
- Maps for the surface potential, $U_{\text{CPD}}$ and work function $\Phi$;
- Mean values and standard deviation of surface potential and work function at the positions defined by the sampling plan.

# 2  Measurement of sheet resistance by Microwave Resonant Cavity

The microwave dielectric resonator method is a non-contact technique to measure the sheet resistance, $R_S$, of monolayer and few layers graphene. It can provide a fast average sheet resistance measurement over a large-area sample without the need of detailed modelling.

The measurement protocol in this section mainly focuses on a closed cavity setup operated in a transmission configuration [1, 2], but various variations of this setup developed at NPL exist in the literature. These include, for example, a similar closed cavity with dielectric resonator operating in reflection configuration [2], and an open cavity setup which can allow non-contact mapping of sheet resistance [2-4] over large areas of graphene. A further modification to this technique using a different resonance mode and a low dc magnetic field has been demonstrated to be capable of non-contact measurements of charge carrier density and mobility in addition to only sheet resistance measurements [5].

It is also worth noting that another non-contact method for sheet resistance measurement using microwave resonant cavity with a different setup has been developed by NIST [6] and has been incorporated into an IEC standard [7].

## 2.1  General

### 2.1.1  Measurement principle

The measurement principle is based on the perturbation of a high-Q dielectric resonator due to the presence of a graphene sample close to the microwave dielectric resonator.

When a graphene sample on a plain low-loss dielectric substrate, of thickness $t_s$ with known real part of the substrate permittivity $\varepsilon'_s$, is brought to a near-field position of a high-Q dielectric resonator, a shift in both the resonant frequency $\Delta f_g$ and the linewidth $\Delta w_g$ can be observed, with respect to the unperturbed resonant frequency $f_0$ and linewidth $w_0$ of a typical resonance mode (TE mode only) of the dielectric resonator. Similarly, the resonance frequency shift $\Delta f_s$ and linewidth shift $\Delta w_s$ when placing only the bare substrate of the same size as the sample at the same position can also be measured. The sheet resistance of the graphene layer can then be calculated from the above parameters as,

$$R_S = \frac{\Delta f_s}{\pi f_0 \varepsilon_0 (\Delta w_g - \Delta w_s)(\varepsilon'_s - 1) t_s} \quad (2.1)$$

More detailed theoretical description of the measurement principle can be found in Ref. [1-3].

### 2.1.2  Sample preparation method

The samples that are suitable for this measurement are uniform monolayer and few layers graphene of reasonably high quality without too high level of macroscopic defects. The graphene layer needs to be situated on one side of an insulating substrate with low microwave loss. For CVD grown graphene, this can be done by transferring the graphene layer onto a clean and thin quartz substrate. For graphene layers that are epitaxially grown on SiC, the SiC substrate can be used directly, provided that only one side of the SiC substrate is covered by graphene, i.e., for samples with graphene grown on both faces, graphene layer on the unwanted face should be etched away before preforming this measurement.

The recommended size of test specimen is from 3 mm × 3 mm to 12 mm × 12 mm. The substrate should be thin, typically 200 µm to 500 µm thick, such that the electric field inside the microwave cavity can be considered as uniform throughout the thickness.

A second piece of identical substrate with the same dimensions but without graphene on top should also be prepared.

Ideally, the sample should have a full coverage of a uniform layer of graphene on top of the substrate. Necessary checks should be performed to ensure the surface area of graphene $A_\text{g}$ is equal to that of the substrate $A_\text{s}$ such that Eq. 2.1 can be used. If the graphene layer only partially covers the low-loss dielectric substrate, the above calculated sheet resistance in Eq. 2.1 should be multiplied by a factor $C = A_\text{g}/A_\text{s}$, corresponding to the graphene coverage.

### 2.1.3 Description of measurement equipment / apparatus

The measurement setup for the measurement of $R_\text{s}$ with the microwave resonance cavity method consists of:

- A cylindrical copper housing;
- A cylindrical single crystal sapphire puck;
- Multiple supporting quartz tube spacers;
- An automatic two-port vector network analyser (VNA) operating in the frequency range 3 GHz to 12 GHz with the capability of measuring the scattering wave parameters $S_{21}$ or $S_{12}$ transmitted between ports 1 and 2 of the VNA;
- High quality coaxial cables and appropriate adapters;
- A data acquisition and processing system equipped with an IEEE 488.3 I/O interface or equivalent;
- An optical microscope (optional).

The cylindrical copper housing is used as a resonance chamber, typically 20 mm in inner diameter and 20 mm in height. Microwave power is coupled in and out of the housing through SMA connectors, passing through the cylindrical wall of the housing at its mid-height and terminated in L-shaped copper wire antennas, oriented with one end soldered to the connector centre conductor, the other arm of the L being directed along the circumference of the housing interior. This orientation ensures that only TE modes in the dielectric resonator will be excited.

The cylindrical single crystal sapphire puck is used as a dielectric resonator. The sapphire puck has relative permittivity of 11.6 in the *c*-axis and 9.4 in the *a-b* plane. It typically has dimensions of 12 mm in diameter and 3 to 5 mm in height. It is situated inside the copper housing and supported by a short quartz spacer tube to separate it from the copper housing. The copper housing, the sapphire resonator and the quartz spacer tube are all aligned to a common axis.

Quartz tubes spacers (or other low-loss low permittivity cylinders such as rigid polymer foam) with various lengths are used to support both the sapphire resonator and the sample/substrate during measurement. The length should be adjusted (usually a few mm) according to the sample conductivity to achieve an optimal level of resonator perturbation.

The vector network analyser (VNA) is used to measure the S-parameters of a two-port resonator. The operating frequencies of a typical VNA (Hewlett Packard 8720) can range from 2 to 20 GHz. The bandwidth measurement function of this instrument allows the transmission $S_{12}$ parameter as a function of frequency to be measured.

The data acquisition and processing system consists of a computer which downloads the data that collected by the VNA, and a LabVIEW programme to perform a non-linear least-squares fitting routine to a skewed Lorentzian lineshape to determine the resonant frequency and the linewidth.

The optical microscope or similar optical magnifying apparatus is an optional piece of equipment to check the uniformity of the graphene sample if partial graphene coverage is suspected. It can be connected a computer to capture and analyse the image of the sample surface to estimate the graphene coverage.

A sketch of a typical setup is shown in the figure below.

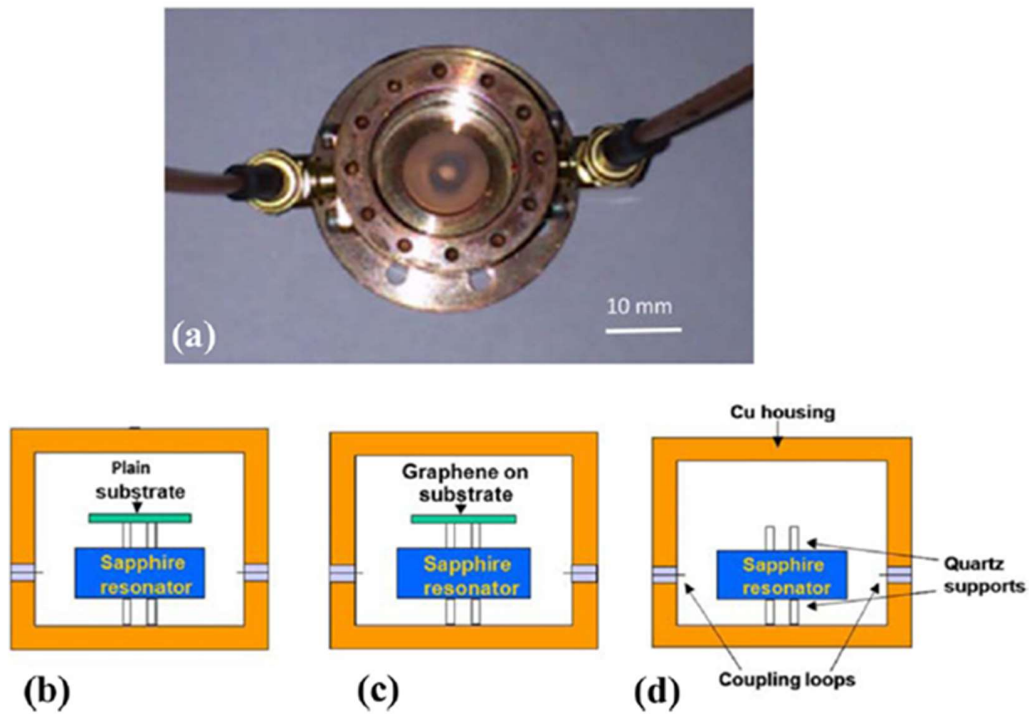

**Figure 2.1: Typical setup for the microwave resonant cavity method**

Figure 2.1 shows: (a) Photo of sapphire puck and quartz spacer tubes inside copper housing (lid removed). (b)-(d) Schematic diagram of the high-Q sapphire dielectric resonator for measurement of the surface impedance of graphene samples. (b) A plain substrate, (c) an identical substrate with graphene film, and (d) with neither substrate, just the dielectric resonator and support structures. Adapted from Ref. [1].

### 2.1.4 Ambient conditions during measurement

Being a carbon monolayer, graphene electrical properties are strongly affected by ambient conditions, in particular the humidity [8,9]. The ambient conditions during the whole duration of the measurement must be monitored and recorded. Typical ambient conditions are those of electrical calibration laboratories, T = (23 ± 1) °C, RH = (50 ± 4) % but can be chosen from the table 1 in of IEC 60068-1.

## 2.2 Measurement procedure

### 2.2.1 Calibration of measurement equipment

VNA calibration should be performed, e.g. according to Guidelines on the Evaluation of Vector Network Analysers (VNA) published by EURAMET [10].

### 2.2.2 Detailed protocol of the measurement procedure

Resonance peaks of the combined housing and the dielectric resonator are first determined by measuring the transmission $S_{12}$ parameter across the whole frequency range, typically between 2 to 20 GHz. The resonance modes acquired are predominantly in the sapphire region which can be distinguished from the other modes corresponding to the empty region of the copper housing and mixed modes in both regions, by the observed resonance $Q$ factor. The resonance modes in the sapphire region have significantly higher quality factor (typically around $10^4$) whereas the other modes have $Q$ factors about an order of magnitude lower. Only TE modes with electric fields in the *a-b* plane are used for this measurement.

Connect the empty test fixture to the vector network analyser (VNA). Set the VNA to measure $S_{21}$ magnitude with 200 data points or more. Select the frequency span to 500MHz and the centre frequency to 10.6 GHz. A single resonant peak should appear on the VNA screen, with the $|S_{21}|$ max value of about – 20 dB and the $S_{21}$ minimum value (noise floor) in the range of – 60 dB or less. Identify the resonant frequency ($f_0$) of this resonant peak $TE_{110}$, for which the electric field of the standing wave is optimally transverse and circularly symmetric.

Set the centre frequency to $f_0$ and the frequency span to about 5$\Delta f$ (10 MHz or less), such that $|S_{21}|$ peak height is about 5 dB. Determine the half power bandwidth, $I = |f_2 – f_1|$, where $f_2$ and $f_1$ are frequencies of the resonant peak, 3dB below the $|S_{21}|$ maximum. Determine the quality factor $Q_0$ of the empty cavity from equation the $Q = f/\Delta f$.

A bare substrate is brought to the vicinity of the sapphire puck and placed on top of the puck supported by the quartz tube spacer. Align the centre of the substrate to the central axis of the system. The resulting frequency $f_s$ and linewidth $w_s$ are measured again. A reasonable choice of the substrate type and thickness can make sure there is no significant reduction of the $Q$ factor.

Remove the bare substrate and replace it with graphene sample, which is graphene layer covering the nominally identical substrate. The graphene sample should be placed at the same position as the bare substrate. Measure again the resulting frequency $f_{g1}$ and linewidth $w_{g1}$ with the graphene layer facing up and the frequency $f_{g2}$ and linewidth $w_{g2}$ with the graphene layer facing down. Adjust the centre frequency and the frequency span if necessary.

Record the ambient conditions. It is well known that the electrical properties of graphene can be influenced by the ambient temperature and humidity. In addition, the dielectric resonator frequency can be changed if the ambient temperature changes throughout the measurement procedure. Therefore, it is advisable to measure the temperature and humidity before and after the measurement is carried out.

The whole measurement procedure normally takes less than 2-3 mins.

### 2.2.3 Measurement accuracy

The uncertainty due to random errors are relatively small when the measurement is performed with careful choice of substrate thickness and spacer height, giving high Q factor and low SNR. The measurement accuracy can be further improved by typically ten times using the above-mentioned LabVIEW program to perform the fitting routine than only by repeated measurements. The reproducibility of this method between measurement to measurement is very high with typical random uncertainty of less than 1%.

On the other hand, the main attribution to the measurement uncertainty for this method is due to multiple approximations and assumptions in the theoretical models to derive the sheet resistance [1-3]. Firstly, strictly speaking, the conductivity or resistivity of graphene is frequency dependent and complex valued. However, in the microwave regime with frequency below at least about 20 GHz, the frequency dependence is negligible with nearly zero imaginary part according to Ref. [11, 12]. The microwave conductivity has therefore been considered to be equal to the dc conductivity. Secondly, it is assumed that the change of the electromagnetic field distribution due to graphene at microwave frequencies is small because the skin depth for monolayer graphene is far greater than its thickness. It is also assumed that the electric field is uniform across the thickness of both the substrate and graphene. This can be confirmed if the deduced sheet resistance is the same for both orientations (graphene facing up and down). In practice, however, it is found that the sheet resistance differences between both orientations can often be as large as 10%. Therefore, averaging the deduced values for both orientations is necessary for this method, and significantly improves the measurement accuracy. Thirdly, the deduced sheet resistance is an average value over the whole area of graphene coating, weighted according to the transverse electric field intensity at various positions of the graphene sample surface. The weighting is usually not uniform across the surface depending on the exact TE mode used. Averaging the deduced sheet resistance with different displacements of the sample can yield a more accurate result.

It is noted that the measurement uncertainty also depends on the absolute value of the graphene sheet resistance which is subjected to the level doping and defects in the graphene layer. By optimising the spacer tube height, sheet resistances of graphene at least in the range from 10 to $10^6$ Ω/sq can be accurately measured [2].

### 2.2.4 Data analysis / interpretation of results

The frequency and the linewidth shift due to the bare substrate compared to the unperturbed values are,

$$\Delta f_s = |f_s - f_0| \qquad (2.2)$$

$$\Delta w_s = |w_s - w_0| \qquad (2.3)$$

The averaged frequency and the linewidth shift due to the graphene sample are calculated as,

$$\Delta f_g = \left| \frac{(f_{g1}-f_0)+(f_{g2}-f_0)}{2} \right| \quad (2.4)$$

$$\Delta w_g = \left| \frac{(w_{g1}-w_0)+(w_{g2}-w_0)}{2} \right| \quad (2.5)$$

The sheet resistance $R_S$ of the measured graphene sample can then be calculated using Eq. 2.1 in Section 2.1.1.

## 2.3 Results to be reported

### 2.3.1 General

The results of the measurement shall be documented in a measurement report, including the date and time of the measurement as well as the name and signature of the person responsible for the accuracy of the report.

### 2.3.2 Product / sample identification

The report shall contain all information to trace back the history of the sample, such as:

- Supplier;
- Trade name;
- Traceability method: Batch number, serial number, etc;
- ID Number;
- Manufacturing date;
- Specification;
- Material type;
- Substrate;
- Sample geometry: Technical drawing or similar.

### 2.3.3 Test conditions

The laboratory ambient conditions during the test:

- Atmosphere;
- Temperature range: 22 °C < T < 24 °C;
- Range of relative humidity: 45 % < RH < 55 %.

### 2.3.4 Measurement specific information

Documentation on the applied setup should comprise information on the applied measurement device (type, manufacturer). Especially:

- Calibration status for the VNA used;
- Sample dimensions (including substrate of the layer to be investigated);
- Used resonance modes and frequencies of the unperturbed cavity;
- Sampling plan;
- Substrate used in the measured (can be different from the substrate which the sample was grown on) and the substrate's thickness and relative permittivity;
- Height of the spacer tube(s);
- Sheet resistance range;
- Number of measurements and their respective orientations and sample displacements;
- Typical measurement curve to visualize measurement accuracy and signal-to-noise ratio.

### 2.3.5 Test results

The test results shall be documented and comprise the following information:

- Date and time of the measurement;
- Maps for sheet resistance and/or sheet conductance;
- Mean values and standard deviation of the relevant parameter (sheet resistance and/or sheet conductance) at the positions defined by the sampling plan.

# 3 Measurement of sheet resistance by terahertz time-domain spectroscopy

This part of the document establishes a standardized method to determine the sheet resistance of graphene by terahertz time-domain spectroscopy.

Terahertz time-domain spectroscopy (THz-TDS) is a contactless and non-destructive method to measure electrical properties of graphene and other 2D materials, obtaining single point measurements of the materials or spatial resistance maps along large areas in a quick way, allowing the characterization of the electrical properties distribution and homogeneity [1, 2].

## 3.1 General

### 3.1.1 Measurement principle

The method to measure the intrinsic resistivity of a thin conducting layer as described in this good practice guide is based on the terahertz time-domain spectroscopy (THz-TDS). In this technique, an electromagnetic pulse with a typical duration of 1 ps is generated by a nonlinear interaction between a high energy femtosecond laser pulse and a photoconductive antenna (PCA) or a nonlinear electro-optical crystal (EOX). Figure 3.1 shows a typical THz waveform. The Fourier transform of the time-domain THz waveform provides access to the frequency components of the pulse, which typically spans from 100 GHz to 5 THz.

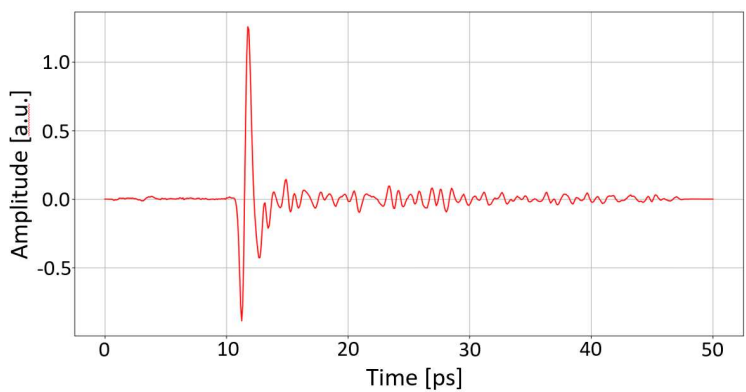

**Figure 3.1: Typical waveform E(t) of the electrical field generated by the femtosecond light pulse.**

This THz pulse is focused on a substrate covered with a thin layer of conducting material such as graphene or another 2D material. The interaction of the incidence THz pulse with the free electrons in the conducting layer modifies amplitude and phase of the transmitted and reflected THz waveform according to the complex refractive index of the material. Assuming that the behaviour of the free electrons can be described by the Drude model, the complex refractive index is related to the intrinsic complex resistivity of the material.

The sheet resistance ($R_s$) is an extrinsic property of a device and the result of an electrical measurement. The sheet resistivity ($\rho_s$) is an intrinsic, local property of a material, and is the 2D equivalent of 3D resistivity. While for many bulk 3D materials, such as pure metals, a consistent value for the resistivity can be given, this is not generally the case for 2D materials, as these are more vulnerable to environmental conditions, in contact with foreign objects and surfaces, and very frequently contain tears, holes, gaps, cracks and other flaws that affect the current flow.

THz-TDS technique allows measuring the following electrical parameters:
- sheet resistance ($R_s$)
- sheet resistivity ($\rho_s$)
- sheet conductance ($G_s = 1/R_s$)
- sheet conductivity ($\sigma_s = 1/\rho_s$)
- mobility ($\mu$)
- carrier density ($N$)
- substrate refractive index ($n$).

This good practice guide describes how to measure the sheet resistance. However, the signal information recorded can be processed to obtain the other electrical parameters listed above.

Sheet resistivity, sheet resistance and mobility are only defined for diffusive conductors, i.e. materials that have lateral (parallel to the plane) dimensions significantly larger than the relevant transport lengths, such as the mean free path (momentum relaxation length) and the de Broglie (Fermi) wavelength ($\lambda_F$). The definitions in this document apply only to conductors which are diffusive, i.e. graphene films and other 2D electric materials.

There are two THz-TDS measurement modes: single point mode and imaging mode.

- Single-point mode refers to a basic THz-TDS measurement where the THz beam is pointed at a relevant position on the surface of the sample. Then, the sample signal is recorded and compared to a reference measurement to obtain the resistivity at that point with a resolution given by the spot size. The resolution of the equipment can be determined by measuring the spot size of THz beam on the sample by the knife-edge, for example.
- Imaging mode refers to raster scanning of a relevant area on the sample, where sample signals are recorded at each pixel position and compared to a reference measurement. In this manner, spatial resistance and conductivity maps of the relevant area can be measured [3, 4, 5].

Mapping of the surface of graphene or another 2D material is required to study the distribution of its electrical parameters ($\rho_s$, $\sigma_s$, $N$ and $\mu$) along the surface. The statistical distributions of the carrier density and mobility are primary descriptors of the sample electrical uniformity. The same statistical distribution can, however, lead to great variations in the current flow, the measured sheet resistance and carrier mobility, depending on their spatial distribution. For such typical parameters, and in contrast to electrical measurements, the THz-TDS method is measuring the intrinsic (local) sheet resistivity $\rho_s(x,y)$ averaged over an area corresponding to the THz spot size.

THz-TDS can be performed in two measurement configurations: transmission and reflection configurations [6].

In transmission geometry, the THz emitter (Tx) and receiver (Rx) are placed in front of each other along the same optical axis and the sample is placed between them. Transmission measurements require the substrate to be transmissive to THz radiation, which is the case for quartz, fused silica, undoped silicon, and various other dielectric and low-doped semiconductors. The frequency-dependent dielectric properties (absorption and index of refraction) of the substrate must be known for a quantitative determination of the resistivity of the graphene layer. These properties can be measured precisely by THz-TDS in a separate reference measurement.

In reflection geometry, emitter and receiver are placed on the same side of the sample. Reflection geometry is convenient to implement, for instance, in the case of absorbing substrates or environments where transmission measurements may not be possible. Practical implementation of the system can be more compact than in transmission geometry.

### 3.1.2 Sample preparation method

No sample preparation is needed to measure graphene samples by THz-TDS. Samples must be placed on the sample holder with the graphene layer upwards.

### 3.1.3 Description of measurement equipment / apparatus

**Principal components of a THz-TDS system**

The principal components of a standard THz-TDS system are listed below and schematized in Figure 3.2:

- Femtosecond laser (fs LASER).
- Beam splitter (BS): it splits the laser output into beams used for THz generation and THz detection.
- Time delay stage: one of the arms of the beam splitter contains an optical delay line for the time scan of the THz transient.
- THz emitter: it generates ultrashort THz pulses when excited by the femtosecond laser. It is typically in the form of a photoconductive antenna (PCA) or a nonlinear electro-optical crystal (EOX).
- THz detector: it is gated by the second part of the laser beam for time-resolved detection of the THz transient. It is typically another PCA or EOX. The detection signal (photocurrent from the PCA or phase retardation of a probe laser beam through the EOX) is detected by e.g. lock-in techniques and recorded as a function of the position of the optical time delay.

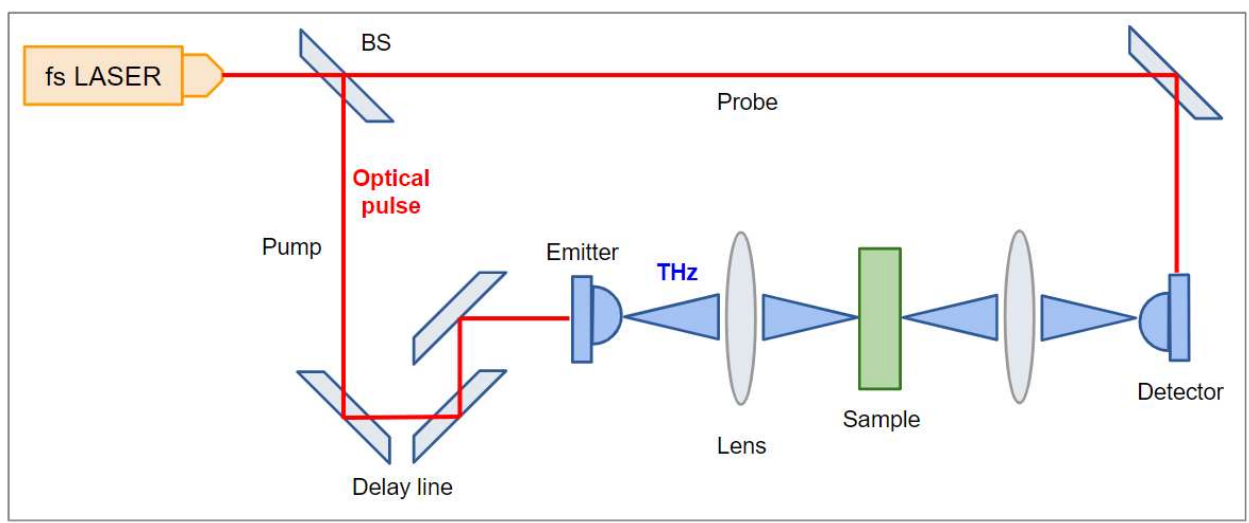

**Figure 3.2: Principal components of a standard THz-TDS system.**

Other THz sources different from femtosecond laser-based sources, can be also used in a THz-TDS system (e.g. electronic generation by microwave up-mixing).

THz-TDS systems rely on ultrashort, broadband THz pulses, so the methodology is not compatible with the use of continuous-wave, narrowband THz sources and detectors. Imaging or mapping is performed by raster scanning of the sample in the focal plane and recording of the THz transient as function of the position (x, y) of the sample. Complete THz-TDS systems are available from several commercial suppliers.

**Measurement configurations**

Two different measurement configurations are possible.

In transmission geometry, the THz emitter (Tx) and receiver (Rx) are placed in front of each other along the same optical axis. The sample is placed between them. The THz beam can be either collimated or focused by reflective or refractive optical elements. Figure 3.3 shows diagrams of both collimated and focused transmission geometries. When the beam is focused, the sample is often placed at the focal plane. Standard mathematical analysis of the signals recorded in the experiment assumes plane-wave excitation of the sample, so it is of importance that the sample is thin compared to the Rayleigh range $z_R$ in order to meet this approximation. In this context, thin substrates fulfil $nd \ll z_R$, where $n$ is the refractive index of the substrate and $d$ is the thickness of the substrate.

The optics for focusing or collimating the THz beam can be either refractive (lenses) as indicated in Figure 3.3 (a, b) or reflective (parabolic or elliptical mirrors) as indicated in Figure 3.3 (c, d).

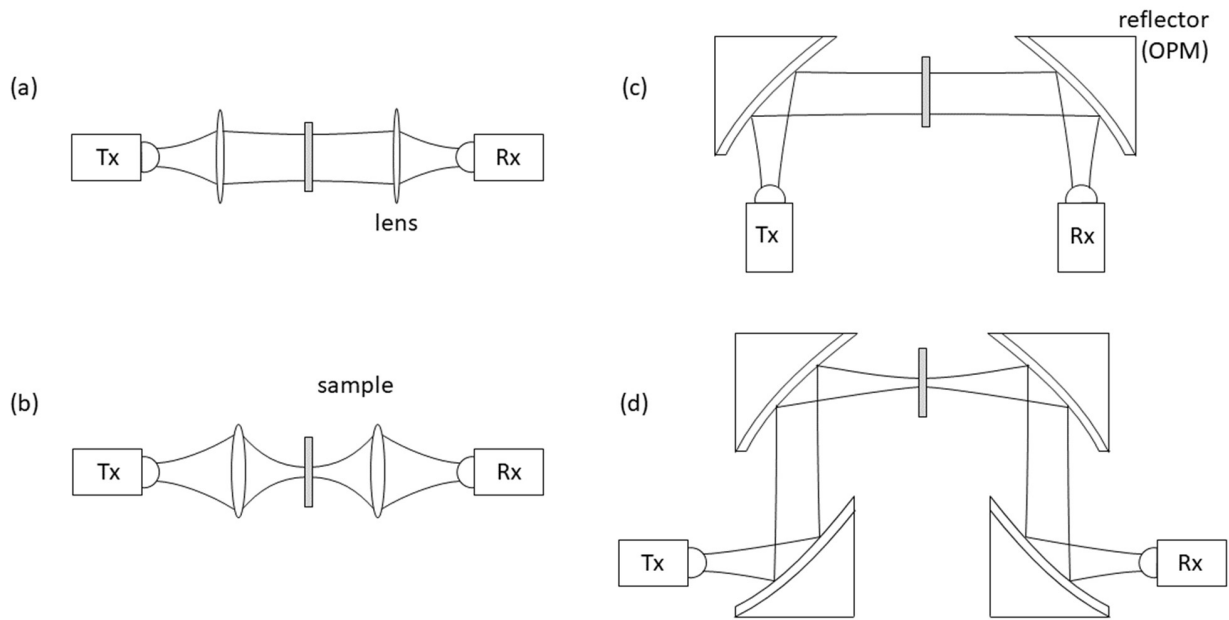

**Figure 3.3: Comparison of different measurement geometries. Collimated (a) and focused (b) transmission geometry with refractive optical elements. Collimated (c) and focused (d) transmission geometry with reflective optical elements.**

In reflection geometry, emitter and receiver are placed on the same side of the sample. The procedure to extract the parameters is more complex than in transmission geometry, and in systems with a focused THz beam it is much more sensitive to the exact positioning of the sample with respect to the focal plane. There are two possible reflection geometries: shallow angle (or pitch-catch) and normal-incidence reflection (Figure 3.4). In pitch-catch, the THz beam illuminates the sample at an oblique angle; the specular reflection is then captured and measured by the detector (Figure 3.4(a)). In normal-incidence reflection, the THz beam illuminates the sample perpendicular to the surface; the incident and reflected beams are separated by a beam splitter (Figure 3.4(b)).

The advantages of normal incidence compared to pitch-catch is a tighter focal spot and better beam quality. In the pitch-catch geometry, the spot size on the sample in the plane of reflection is enlarged by the factor $1/\cos\theta$ where $\theta$ is the incidence angle. However, the beam splitter introduces a minimum loss of 6 dB. Therefore, the pitch-catch geometry offers a higher signal because losses are minimal, but the focal spot is larger, so spatial resolution is worse than in normal reflection. Both geometries are similar from the parameter extraction point of view.

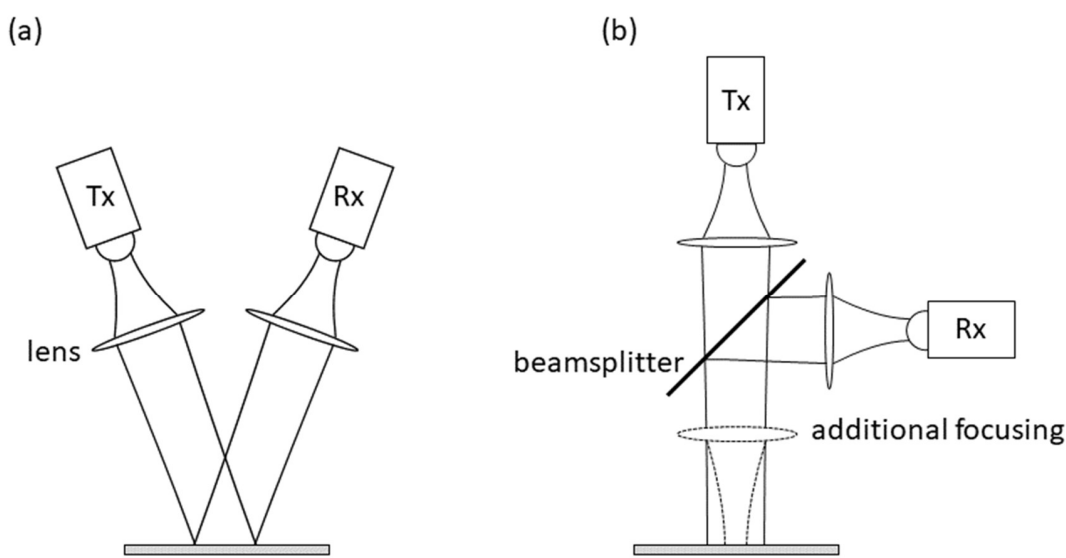

**Figure 3.4: (a) Pitch-catch reflection geometry. (b) Normal-incidence reflection geometry.**

### 3.1.4 Supporting materials

Reference signals are needed to calculate the resistance of the sample. Measurement of the reference signal depends on the measurement configuration:

- For reflection measurements, reference signals can be measured on the bare substrate (at a position on the sample without graphene) or on a reference material (e.g. metal).
- For transmission measurements, reference signals can be measured through a bare substrate (at a position on the sample without graphene; the bare substrate must be transmissive to THz radiation) or without sample in the beam path.

Several considerations related to the substrates are relevant to obtain reproducible measurement results. They are explained below.

- **Substrate dielectric properties (transmission + reflection)**

    The frequency-dependent index of refraction and the absorption coefficient of the substrate must be known in order to calculate the resistivity of the graphene on top of the substrate. For many standard substrates, this information is available in the open literature. However, the THz-TDS

methodology is still lacking standardised measurement protocols and published literature values cannot be traced to international standards. Thus, significant discrepancies between independent measurements exist. It is therefore recommendable that a separate THz-TDS characterization of the substrate is performed. For maximum consistency, the same frequency resolution and bandwidth as for the subsequent graphene conductivity measurements shall be used.

- **Thickness of the substrate (transmission)**

  Transmission measurements of the full complex-valued thin-film conductivity are highly sensitive to the precise thickness of the substrate. Although the average substrate thickness is unimportant while the substrate is transmissive to THz radiation, any unintentional thickness variations (even in the sub-micrometer range) can influence the phase difference between sample and reference, and therefore should be known [7,8]. It is customary to use substrate thicknesses in the range 0.1-1 mm.

- **Reflection plane of the sample (reflection)**

  Reflection measurements of full complex-valued thin-film conductivity are highly sensitive to the precise location of the reflection plane of the sample along the propagation direction of the THz beam. The error on the frequency-dependent phase of the reflected signal is given by $\delta\phi=4\pi\nu\Delta z/c$. The numerical value of this phase error is 0.042 mrad/THz/nm. For 100 nm positioning error at 1 THz, the phase error is thus 4.2 mrad. This error is particularly difficult to avoid if separate materials are used for sample and reference measurement (for instance a smooth metal reflector for reference measurements), but also measurements with sample and reference points on the same substrate can have positioning errors that become significant.

### 3.1.5 Calibration standards

To check the correct performance of the equipment, two calibration standards can be used:

- For transmission measurements: air.
- For reflection measurements: a highly reflective surface, such as a mirror of 52-54 HRc tempered steel.

The result of measuring air in transmission or the steel mirror in reflection should be constant along the time if the equipment works properly.

### 3.1.6 Ambient conditions during measurement

- Temperature range:        20 °C < T < 35 °C.
- Range of relative humidity:   RH < 70%.
- Atmosphere:               air, nitrogen, argon (at atmospheric pressure).

Thermal equilibration of the measurement setup and the sample itself must be assured before the measurement. Typically, at least one hour of thermal equilibration is desirable after powering up the measurement system.

The measurements shall be performed at constant temperature and humidity so that consecutive measurements of the same or different samples are comparable.

It should be noted that the humidity of the lab might have a significant effect on the graphene conductivity even if the measurements are performed in a dry nitrogen atmosphere, unless a thermal treatment in nitrogen is performed at 225 °C to remove surface water [9].

## 3.2 Measurement procedure

This section describes the procedure to perform THz-TDS measurements of sheet resistance, as well as the main factors that affect the accuracy of the measurements. The protocol explained here has been tested against other contact and non-contact methods for the electrical characterization of graphene [10].

### 3.2.1 Calibration of measurement equipment

Calibration of the time axis is necessary to ensure a correct frequency axis. The frequency step size $\Delta f$ is determined by the temporal recording range $T$ as $\Delta f = 1/T$. Commercial THz-TDS systems have a factory-calibrated time axis, and thus a calibrated frequency axis. Custom-built THz-TDS systems can be calibrated by recording of a THz transient propagated through the empty spectrometer, in ambient air. The measured spectral position of sharp and characteristic absorption lines from rotational transitions in water vapour molecules in the beam path can be matched up against tabulated reference values by fine adjustment of $\Delta f$, and hence be used for precise calibration of the time axis. Modern motorized translation stages are so precise that this calibration procedure is rarely needed. Relevant information regarding the calibration of the time and frequency axis shall be given in the measurement report.

In order to check the correct performance of the equipment, a measurement of air (transmission), a mirror of 52-54 HRC tempered steel (reflection) or a blank substrate with known refractive index and absorption (reflection and transmission) should be done and compared previous measurements of the same materials.

### 3.2.2 Detailed protocol of the measurement procedure

There are two consecutive steps are, whose order is indistinctive:

- Measure the sample.
- Measure the reference.

By comparing both measurements, it is possible to extract the complex index of refraction of the sample and, thus, extract the reflectance. The path between the emitter and the receiver can be purged with dry air, nitrogen or vacuum to reduce or suppress the effects of water vapour adsorption. In both cases, the time traces of the electric field, $E_{ref}(t)$ and $E_{sam}(t)$, shall be recorded.

It shall be ensured that the THz beam is not disturbed by the edges of the sample or any metallic structures (e.g. electrodes or contact pads). Thus, the free sample area must be large enough so that the THz-TDS signal in contact with any metal contacts or other device features do not affect the measurement. As a guideline, this is the case 1.2 mm away from any edges or metallization on the sample at 0.9 THz [11], and data closer to the contacts can be neglected. Depending on the choice of step-size used during the THz-TDS experiment, samples with dimensions of several millimeters are large enough to obtain hundreds of resistivity pixels.

**Transmission configuration**

Reference signals can be measured through a bare substrate (at a position on the sample without graphene; the bare substrate must be transmissive to THz radiation) or without sample in the beam path.

The measurement protocol is described as follows:

1. In case of a purged measurement, wait until the humidity level in the measurement chamber is <1%.
2. Set a temporal scan range of the THz system to a fixed range that should be wide enough to contain the relevant information. The start of the scan window should be so early with respect to the THz signal shape that the signal has decayed below the noise floor. The end of the scan window is defined by the substrate thickness ($d_{sub}$), since the first time-domain echo ($t_{echo}$) from the substrate will be at $t_{echo} = n_{sub}d_{sub}/c$ (where $n_{sub}$ is the substrate refraction index and $c$ is the speed of light in a vacuum). Longer windows can be used if the time-domain echoes are included in the data extraction algorithm.
3. Acquire a waveform without any sample. This waveform is the reference measurement.
4. Block the THz path with a metallic object.
5. Acquire a waveform with the THz path blocked. This is a measurement of the noise floor.
6. Unblock the THz path and place the sample at the measurement location.
7. In case of a purged measurement, check that humidity level is still <1%.
8. Acquire a waveform of the sample in the same conditions as with the reference. This waveform is the sample measurement.
9. Extract the electrical parameters of the sample.

Steps 1 and 7 are not strictly necessary for every measurement and may be omitted; reducing the humidity level mitigates the effects of atmospheric water vapour absorption in the measurement, which introduces narrow resonances in the spectrum.

**Reflection configuration**

Reference signals can be measured on the bare substrate (at a position on the sample without graphene) or on a reference material (e.g. metal).

The measurement protocol is described as follows:

1. In case of a purged measurement, wait until the humidity level in the measurement chamber is <1%.
2. Set a temporal scan range of the THz system to a fixed range that should be wide enough to contain the relevant information. The start of the scan window should be so early with respect to the THz signal shape that the signal has decayed below the noise floor. The end of the scan window is defined by the substrate thickness ($d_{sub}$), since the first time-domain echo ($t_{echo}$) from the substrate will be at $t_{echo} = 2n_{sub}d_{sub}/c$ (where $n_{sub}$ is the substrate refraction index and $c$ is the speed of light in a vacuum). Longer windows can be used if the time-domain echoes are included in the data extraction algorithm.
3. Acquire a waveform of a perfect reflector surface (e.g. a mirror or optically polished metal). This waveform is the reference measurement.
4. Block the THz path with a metallic object.
5. Acquire a waveform with the THz path blocked. This is a measurement of the noise floor.
6. Unblock the THz path and place the sample at the measurement location.

7. In case of a purged measurement, check that humidity level is still <1%.
8. Acquire a waveform of the sample in the same conditions as with the reference. This waveform is the sample measurement.
9. Extract the electrical parameters of the sample.

Steps 1 and 7 are not strictly necessary for every measurement and may be omitted; reducing the humidity level mitigates the effects of atmospheric water vapour absorption in the measurement, which introduces narrow resonances in the spectrum.

With minor modifications of the data extraction methodology, step 3 can also be performed using a dielectric surface (e.g. a blank section of the sample substrate). In this case, the (known) reflectivity of the blank dielectric ($r_{dielec} = 2n/(n+1)$) must be used instead of the unity reflectivity of a metal surface ($r_{metal} = -1$).

### 3.2.3 Measurement accuracy

The accuracy of the measurement depends of many factors:

– Thickness variations in the sample. The time shift of the pulse is directly proportional to the optical thickness of the sample, which is the product of the real thickness and the index of refraction. Therefore, variations in the thickness of the sample will cause a linear change in the variations in the waveform time-shift.
– Phase variations due to the pulse jitter. Some systems show a change in the apparent position of the pulse caused by jitter in the laser or in the delay line [12].
– Time-domain waveform windowing. Frequency information is obtained through a Fourier transform of the waveform. In order to clean up the signal before the Fourier transform, a temporal window cuts away unwanted parts of the full recorded waveform. Different mathematical forms of window functions applied to the waveform will affect the Fourier transform. In particular, the Fourier transform of the window will be convolved with the Fourier transform of the waveform itself. It is known that some window functions tend to affect the magnitude of the peak while enhancing the frequency resolution, whereas some windows tend to preserve the magnitude of the peaks while losing some frequency resolution. The application of different types of windows is often determined by the type of measurement. A test for correct performance of signal windowing shall be performed to verify that a measurement that applies windowing can reproduce the refractive index of e.g. high-resistivity silicon, yielding a constant refractive index of 3.417 and absorption coefficient below 0.05 cm$^{-1}$ for frequencies below 2.5 THz [13]. The result of this test shall be given in the measurement report.
– Signal-to-noise-ratio (SNR). The higher the SNR, the more accurate the measurement because under such conditions the contribution of the random noise to the measurement is minimal. The opposite is true, the lower the SNR, the less accurate the measurement because random noise is more noticeable. The SNR shall be given in the test report.

## 3.3 Data analysis / interpretation of the results

The main steps of the data analysis are described as follows:

1. Transform the reference and sample signals (time-domain waveforms) to frequency domain waveforms through a Fourier transform.
2. Divide the sample frequency-domain waveform by the reference frequency-domain waveform to obtain the frequency-dependent amplitude and phase of the ratio.
3. Ensure that the phase of the ratio extrapolates to zero at DC frequency ($f$=0).
4. Use the amplitude and phase of the ratio obtained in step 2 to calculate the resulting frequency dependent complex sheet resistivity of the graphene.
5. If multiple measurements on the same material (or the same position on a graphene surface) are available, the standard deviation of the measurement at each frequency point shall be calculated, and thus a complete measurement (value and uncertainty) shall be reported.
6. Other parameters such as sheet conductivity, carrier density and mobility can be also calculated.

Steps 1-3 are discussed in detail in reference [14].

## 3.4 Results to be reported

### 3.4.1 General

The results of the measurement shall be documented in a measurement report, including the date and time of the measurement as well as the name and signature of the person responsible for the accuracy of the report.

### 3.4.2. Sample identification

The report shall contain all the information to identify the test sample and trace back the history of the sample:

- General procurement information.
- General material description including a technical drawing.
    - Picture of the top view of the sample, indicating the inspected area and location of the measurement positions.
    - Cross section (side view), showing the layered structure.

### 3.4.3 Test conditions

The laboratory ambient conditions during the test must be constant and should be the following:

- Atmosphere: air, nitrogen or argon.
- Temperature range: 15°C < T < 35°C.
- Range of relative humidity: RH < 75%.

### 3.4.4 Measurement specific information

- Measurement configuration: reflection, transmission.
- Propagation characteristics of the beam: focused, collimated.
- Bandwidth of the system (units of THz).
- Signal to noise ratio (units of dB at a given frequency).
- Frequency of the results (units of THz and/or µm).
- Reference material.
- Spot size of the THz beam.
- Sampling plan for single point mode.
- Scanning step width for imaging scanning mode.

### 3.4.5 Test results

The following items shall be part of the test report.

- Sampling plan: coordinate system used in the measurement setup in absolute positions with a definition of the origin so that the measurement locations can be related to the technical drawing of the sample.
- Results of sheet resistance or sheet conductance measured according to this standard: table of mean values and standard deviations of sheet resistance and/or sheet conductance at the positions defined by the sampling plan.
- Graphical visualization of sheet resistance or sheet conductance measured according to this standard: maps for sheet resistance and/or sheet conductance. The colour map shall be scaled in absolute positions in respect of the origin of the coordinate system. The colour code shall be calibrated in absolute values of the measured sheet resistance or sheet conductance.

## Conclusions

Using this guide, users with previous characterisation expertise within industry can more reliably, quantitatively and comparably measure the electrical properties of commercially supplied graphene.

This guide forms the basis for future international standards in this area, specifically, standards currently under development within IEC/TC 113 'Nanotechnology for electrotechnical products and systems' for the characterisation of the electrical properties of graphene. This will lead to the continual improvement in the measurement of the electrical properties of graphene, as well as reveal any reproducibility issues in performing these graphene measurements in different laboratories across the world.

## Acknowledgements

The work is part of the European project "GRACE — Developing electrical characterisation methods for future graphene electronics", code 16NRM01. This project has received funding from the European Metrology Programme for Innovation and Research (EMPIR) programme co-financed by the Participating States and from the European Union's Horizon 2020 research and innovation programme.

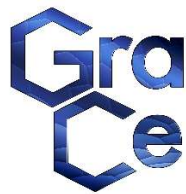